\documentclass[aps,prl,showpacs,twocolumn,groupedaddress]{revtex4}

\usepackage{latexsym}
\usepackage{times}
\usepackage{graphicx}
\usepackage{hyperref}
\usepackage{epsfig}

\begin{document}
\thispagestyle{empty}
\pagestyle{empty}

\bibliographystyle{apsrev}

\title{Manipulating the Loss in Electromagnetic Cloaks for Perfect Wave Absorption}

\author{Christos Argyropoulos}
\email[]{christos.a@elec.qmul.ac.uk}
\affiliation{Department of Electronic Engineering, Queen Mary, University of London, Mile End Road, London, E1 4NS, United Kingdom}
\author{Efthymios Kallos}
\affiliation{Department of Electronic Engineering, Queen Mary, University of London, Mile End Road, London, E1 4NS, United Kingdom}
\author{Yan Zhao}
\affiliation{Department of Electronic Engineering, Queen Mary, University of London, Mile End Road, London, E1 4NS, United Kingdom}
\author{Yang Hao}
\email[]{yang.hao@elec.qmul.ac.uk}
\affiliation{Department of Electronic Engineering, Queen Mary, University of London, Mile End Road, London, E1 4NS, United Kingdom}

\date{\today}

\begin{abstract}
We examine several ways to manipulate the loss in electromagnetic cloaks, based on transformation electromagnetics. It is found that, by utilizing inherent electric and magnetic losses of metamaterials, perfect wave absorption can be achieved based on several popular designs of electromagnetic cloaks. A practical implementation of the absorber, consisting of ten discrete layers of metamaterials, is proposed. The new devices demonstrate super-absorptivity over a moderate wideband range, suitable for both microwave and optical applications. It is corroborated that the device is functional with a subwavelength thickness and, hence, advantageous compared to the conventional absorbers.
\end{abstract}


\maketitle
Metamaterials can be generally defined as a class of ``artificial'' media that exhibit extraordinary electromagnetic properties not found in nature. For example, materials with negative refractive index can be designed \cite{Shelby}, which can theoretically achieve infinite subwavelength resolution \cite{Pendrylens}. Recently, a dispersive electromagnetic cloak was experimentally demonstrated \cite{Schurig} by achieving the required material anisotropy \cite{Pendry}. Such structures, typically constructed from periodically-placed subwavelength unit cells based on highly conductive metals over dielectric substrates, can be analyzed with the effective medium theory \cite{D.R.Smith}. They are defined by complex electromagnetic parameters: the frequency dependent permittivity $\varepsilon(\omega)=\varepsilon_{1}(\omega)+j\varepsilon_{2}(\omega)$ and permeability $\mu(\omega)=\mu_{1}(\omega)+j\mu_{2}(\omega)$. So far, metamaterial research has mainly focused on the real parts ($\varepsilon_{1},\mu_{1}$) of the material parameters in order to design negative-index and cloaking devices. However, the imaginary parts ($\varepsilon_{2},\mu_{2}$) of the material parameters (which characterize the losses of the medium) can also have interesting potential applications, such as the design of more efficient thermal imagers and novel absorbers. In this letter, we take advantage of the overlooked lossy properties of metamaterials, which are widely regarded as a drawback in the design of these exotic devices \cite{Wiltshire}.

Three absorber designs are proposed, by manipulating both electric and magnetic losses of subwavelength cylindrical electromagnetic cloaking coatings. The first two designs utilize the material parameter sets of the matched reduced \cite{Qiu} and ideal \cite{Cummer} electromagnetic cloaks. The third design consists of a more realistic discrete ten-layer  structure, based on an approach presented in \cite{Schurig}, where only one material parameter is radius-dependent and the others have constant and non-dispersive values. The experimentally verified structure in \cite{Schurig} is generally regarded as an imperfect cloaking device; however, as we demonstrate below, with a proper manipulation of material losses a perfect absorber can be constructed. Specifically, by calculating the field patterns and analyzing the scattering coefficients, we demonstrate that these new devices can achieve perfect wave absorption over a narrow bandwidth and moderate performance over a broad bandwidth. Additionally, they are matched to free space without further parameter tuning and they are found to perform significantly better when compared to conventional dielectric absorbers. The effects are analyzed using a radially-dependent dispersive Finite-Difference Time-Domain (FDTD) method and scrutinized with existing analytical solutions of the cloaks \cite{Ruan,Chen}.

The proposed absorbers are based on a different concept from those used in the recently-demonstrated perfect microwave metamaterial \cite{Landy} and terahertz metamaterial \cite{Tao,HuTao} absorbers. The latter devices operate correctly only when there is strong coupling at the appropriate resonant mode of each metamaterial unit cell. As a result, careful tuning of the complex parameters in the device is essential, in order to achieve zero backscattering and maximum absorptivity. However, the proposed absorbers are independent from resonant modes \cite{Cummer} and, hence, can operate over a relatively wide frequency range.

\begin{figure}[h]
\centering
\includegraphics[width=9.0cm]{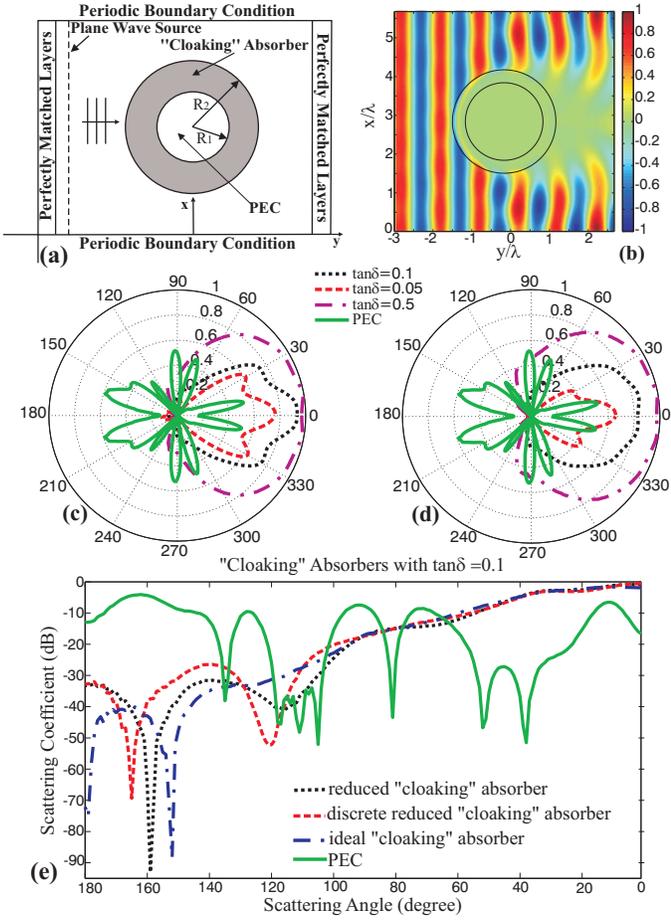}
\caption{(a) 2-D FDTD computation domain of the ``cloaking'' absorber for the case of plane wave incidence. (b) Normalized magnetic field distribution of a subwavelength metamaterial absorber with $\tan\delta=0.5$. The object placed inside the absorption coating is composed of a PEC. (c)-(d) Different scattering coefficient patterns as a function of the loss tangent for the reduced (c) and ideal (d) parameter sets, along with the scattering coefficient pattern of a bare PEC cylinder. (e) Different scattering coefficients of the reduced, ten layer discrete reduced and ideal ``cloaking'' absorbers along with the scattering coefficient of a bare PEC cylinder.} \label{scatteringcoeffabsorber}
\end{figure}

For the two-dimensional (2-D) FDTD simulations presented here, a perfect electric conductor (PEC) cylinder is surrounded by a sub-wavelength absorption coating. Without loss of generality, a TE plane wave is incident and only three field components are non-zero: $E_{x},E_{y}$ and $H_{z}$. The computational domain of the infinite (towards z-direction) cylindrical ``cloaking'' absorber can be seen in Fig. \ref{scatteringcoeffabsorber}(a).

The performance of an absorber can be characterized by its absorptivity $A=1-|S_{11}|^{2}-|S_{21}|^{2}$, where $S_{11}$ and $S_{21}$ are the reflection and transmission coefficients of the device, respectively. An ideal absorber is characterized by an absorptivity equal to unity, which directly leads to no reflection ($S_{11}=0$) and no transmission ($S_{21}=0$) through the device. In addition, the scattering coefficient $\sigma_{S}$ of an absorber can be calculated with reference to free space, with no obstacles present. It is given by the formula $\sigma_{S}=\Big|\frac{|H_{z}|-|H^{fr}_{z}|}{|H^{fr}_{z}|}\Big|$, where $H_{z}$ is the complex magnetic field distribution in the ``cloaking'' absorber and $H^{fr}_{z}$ is the complex magnetic field distribution in the free space.  When $\sigma_{S}=0$, the surrounding field of the absorber is equal to the field in free space, i.e. the structure is totally reflectionless. When $\sigma_{S}=1$, the field is entirely dissipated inside the device ($H_{z}=0$) and the radiation is not transmitted through the absorber. This last condition that $S_{21}\rightarrow0$ in the transmitted region of the absorber ($y>0$ in Fig. \ref{scatteringcoeffabsorber}(a)), combined with the reflectionless property $S_{11}\rightarrow0$ of the cloaking material in the reflected region ($y<0$ in Fig. \ref{scatteringcoeffabsorber}(a)), is the ideal condition in order to achieve a perfect absorber with $A\rightarrow100\%$. The ``cloaking'' absorbing structures proposed in this Letter can easily achieve $A>80\%$, which is rapidly enhanced by increasing the loss factor in some suitable way.

The first design (reduced ``cloaking'' absorber) utilizes the material parameters of the matched (to free space) reduced cloak \cite{Qiu}: $\varepsilon_{r}=\frac{R_{2}}{R_{2}-R_{1}}\left(\frac{r-R_{1}}{r}\right)^{2},\varepsilon_{\phi}=\frac{R_{2}}{R_{2}-R_{1}},\mu_{z}=\frac{R_{2}}{R_{2}-R_{1}}$, where $R_{1}$ and $R_{2}$ are the inner and outer radii of the absorber, respectively, with $R_{1}<r<R_{2}$. The permittivity value is gradually changing with the radius $r$ of the device and it can be obtained that $0\leq\varepsilon_{r}<1$ and $\varepsilon_{\phi},\mu_{z}>1$ for all values of $r$. The $\varepsilon_{r}$ parameter is mapped with the Drude dispersion model: $\varepsilon(\omega)=1-\frac{\omega_{p}^2}{\omega^2-\jmath\omega\gamma}$, where $\omega_{p}$ is the plasma frequency and $\gamma$ is the collision frequency characterizing the losses of the dispersive material. Furthermore, $\varepsilon_{\phi}$ and $\mu_{z}$ are simulated with a conventional dielectric/magnetic model: $\varepsilon(\omega)=\varepsilon+\frac{\sigma}{\jmath\omega}$, where $\sigma$ measures the conductive/magnetic losses. A thorough description of the radially-dependent dispersive FDTD algorithm employed to simulate the proposed absorber can be found in \cite{Yan,Argyropoulos}.

In the case of Drude model mapping of the material parameter ($\varepsilon_{r}<1$), the lossy parameter can be presented in an alternative way: $\hat{\varepsilon}_{r}=\varepsilon_{r}(1-\jmath\tan\delta)$, where the radius-dependent parameter $\varepsilon_{r}$ is given from the reduced parameter set and $\tan\delta$ is the loss tangent of the lossy material. If the previous formula is substituted in the Drude model and $\tan\delta$ is assumed constant, the radially-dependent plasma frequency becomes $\omega_{p}(r)=\sqrt{(1-\varepsilon_{r})\omega^2+\varepsilon_{r}\omega\gamma\tan\delta}$ and the collision frequency becomes $\gamma(r)=\frac{\varepsilon_{r}\omega\tan\delta}{(1-\varepsilon_{r})}$. Similarly, the conductivity of the conventional dielectric/magnetic model is given by (using $\varepsilon_{\phi}$ as an example) $\sigma=\varepsilon_{\phi}\omega\tan\delta$, which is constant because $\varepsilon_{\phi}$ and $\mu_{z}$ have a fixed value. Note that, the concept of artificial loss functions was also introduced in \cite{Zharova}, but for a different coordinate transformation function. It was used to create an absorbing boundary condition (like perfectly matched layers) in computational electromagnetics, whereas the aim of our proposed device is to be implemented in practice for engineering applications.

The device is tested at a frequency of $2$ GHz, which is used throughout the Letter. Other operating frequencies can also be chosen by simply adjusting the material parameters due to the frequency-independent nature of coordinate transformation functions \cite{Pendry}. The dimensions of the device, in terms of the free-space wavelength $\lambda$, are $R_{1}=\lambda$ and $R_{2}=\frac{4\lambda}{3}$. The results of the normalized real part of the magnetic field amplitude distribution when the steady-state is reached are shown in Fig. \ref{scatteringcoeffabsorber}(b), for a device with $\tan\delta=0.5$. The computed scattering coefficients of the device as a function of the loss tangent $\tan\delta$ can be seen in Fig. \ref{scatteringcoeffabsorber}(c), where the scattering coefficient of a bare PEC cylinder is also shown. It is observed that the backscattering of the ``cloaking'' absorber - i.e. angle equal to $180^{\circ}$ - goes to zero for all loss tangents; simultaneously, the reflection coefficients approach zero ($S_{11}\rightarrow0$) as $\tan\delta$ increases. It should also be mentioned that the backscattering coefficient of the reduced ``cloaking'' absorber utilized here is at least $20$ dB lower, when compared to an absorber based on the simple reduced set used in \cite{Schurig} (not shown).

We further verify this absorber concept by introducing a second design which is based on the material parameter set of an ideal electromagnetic cloak \cite{Cummer} (ideal ``cloaking'' absorber), without altering the geometry. All the parameters are now radius-dependent: $\varepsilon_{r}=\frac{r-R_{1}}{r},\varepsilon_{\phi}=\frac{r}{r-R_{1}},\mu_{z}=\left(\frac{R_{2}}{R_{2}-R_{1}}\right)^{2}\frac{r-R_{1}}{r}$. The scattering coefficients of the device, varying with the losses, can be seen in Fig. \ref{scatteringcoeffabsorber}(d). The backscattering of the ideal set device has slightly improved values, compared with the reduced ``cloaking'' absorber in Fig. \ref{scatteringcoeffabsorber}(c). Note that, when the losses are increased, the exact backscattering is still low, but there are weak backward reflections at the sides of both absorbers, which will be discussed later in this Letter.

A big shadow which tends to increase with the loss tangent is cast at the back of both the proposed devices (especially between the angles $-30^{\circ}$ to $30^{\circ}$), as can be seen in Figs. \ref{scatteringcoeffabsorber}(c), (d), which is the ideal scattering pattern for an efficient absorber \cite{andersen2005aer}. It is interesting that, for the low loss tangent of $\tan\delta=0.05$ the absorption of the reduced ``cloaking'' absorber (Fig. \ref{scatteringcoeffabsorber}(c)) is better than the absorption of the ideal ``cloaking'' absorber (Fig. \ref{scatteringcoeffabsorber}(d)). For higher losses ($\tan\delta=0.5$), it is observed that the scattering coefficients tend to one, especially inside the window $-30^{\circ}$ to $30^{\circ}$.

In the previous design both conductivities and collision frequencies are required to be continuously radius-dependent, which makes a practical implementation of this metamaterial structure challenging. In order to check the performance of a more realistic absorber, the device is designed with a discrete ten layered structure (discrete reduced ``cloaking'' absorber), similar to the cloak constructed in \cite{Schurig}. Again the reduced parameter set is used \cite{Qiu} and only the radius-dependent parameter is discretized to ten different values, one for each layer. The other parameters are kept constant, as before, and
the loss tangent is chosen $\tan\delta=0.1$. The scattering pattern of the device is then computed, it is given in Fig. \ref{scatteringcoeffabsorber}(e) and is compared with the continuous reduced and ideal parameter set ``cloaking''
absorbers. It is seen that the patterns of the three absorbers are following a similar trend, especially at the angles between $0^{\circ}$-$90^{\circ}$ (shadow). The larger difference is the backscattering coefficient (angle of $180^{\circ}$), which is less in the case of ideal ``cloaking'' absorber. For $\tan\delta=0.1$, it is calculated that the device achieves an absorptivity of at least
$A=80\%$ within a frequency range of $400$ MHz. If the losses are increased, absorptivity values of $A>90\%$ can be
easily obtained, which suggests that the proposed structure can be utilized as a perfect metamaterial absorber.

\begin{figure}[h]
\centering
\includegraphics[width=9.0cm]{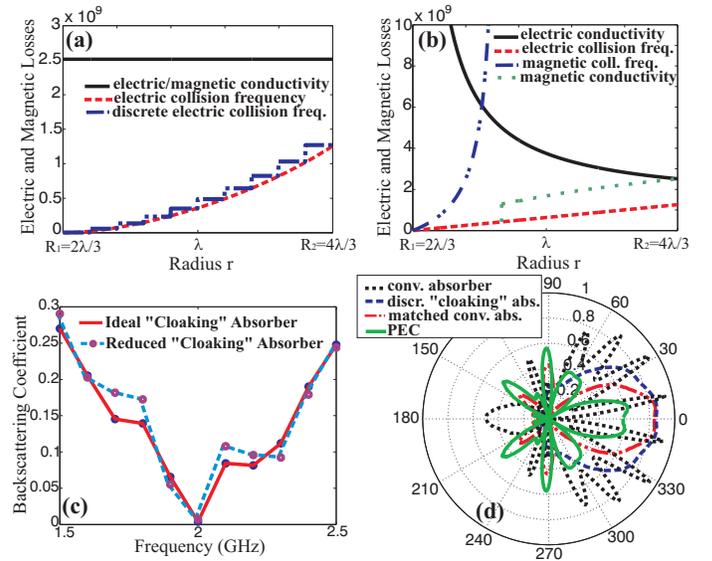}
\caption{(a)-(b) The electric and magnetic losses of the ``cloaking'' absorber as a function of the device's radius for the reduced (a), ten layer discrete reduced (a) and full (b) parameter sets. (c) The frequency-dependent backscattering of the metamaterial absorber with ideal and reduced sets. The bandwidth of the discrete absorber (not shown) is identical to the reduced absorber. (d) The scattering coefficient patterns of a conventional dielectric absorber, a metamaterial discrete reduced ``cloaking'' absorber and a matched conventional absorber ($\tan\delta=0.1$ for all devices), along with the scattering coefficient of a bare PEC cylinder.}\label{comparisonscatteringcoeffabsorber}
\end{figure}

Next, we demonstrate the absorber's performance for different coating thicknesses, with $R_{1}=\frac{2\lambda}{3}$ and $R_{2}=\frac{4\lambda}{3}$. The loss parameters for these structures are graphically depicted in Figs. \ref{comparisonscatteringcoeffabsorber}(a),(b), assuming $\tan\delta=0.1$. Note that at the inner surface of both absorbers ($r=R_{1}$), the electric collision frequency $\gamma_{e}=0$, because $\varepsilon_{r}=0$ at $r=R_{1}$. It can be seen that there are no infinite loss values for the reduced ``cloaking'' absorbers, which makes it easier to be practically implemented.
Moreover, the conductivities of this particular absorber are constant, which leads directly to the usage of only one type of dielectric material as the substrate to fabricate split-ring resonators (SRRs), in order to achieve the required losses. The bandwidth response of the ideal and reduced ``cloaking'' absorbers can be seen in Fig. \ref{comparisonscatteringcoeffabsorber}(c). The lowest backscattering is obtained for both sets at the central frequency of $2$ GHz, while the backscattering is stronger at other frequencies due to the dispersive nature of the absorber. Based on this behavior, the proposed devices can also have moderate wideband applications, e.g. reducing the radar cross section (RCS) of an object.

Next, the ``cloaking'' absorber coating (Fig. \ref{scatteringcoeffabsorber}(a)) is replaced with conventional absorbing materials, in order to evaluate their individual performances. A conventional dielectric absorber with $\varepsilon=2$ and a matched conventional absorber with $\varepsilon=1$ are compared to the discrete reduced ``cloaking'' absorber. All the devices have identical loss tangents of $\tan\delta=0.1$ and the same dimensions ($R_{1}=\frac{2\lambda}{3}$, $R_{2}=\frac{4\lambda}{3}$). The obtained scattering coefficient patterns of these absorption coatings are shown in Fig. \ref{comparisonscatteringcoeffabsorber}(d), where it is observed that the proposed metamaterial device performs better when compared to the conventional ones. Firstly, minimal backscattering is only achieved with the proposed ``cloaking'' structure, which is not the case even for the matched conventional absorber. In addition, the absorption (shadow) of the ``cloaking'' absorber is more uniform and more solid than the other two cases. Furthermore, when compared with the absorption of the matched dielectric absorber, the shadow is notably larger in a broader angle range (between $-90^{\circ}$ to $90^{\circ}$).

One of the main physical reasons that explains the superior performance of the metamaterial absorber designs is the bending of the field wavefronts within the absorption coating. Due to the anisotropy of the material parameters, the impinging field energy is guided over longer distances around the cloaked object, thus, dissipating gradually inside the thin sub-wavelength coating. The proposed `cloaking'' absorbers inherit this unique property from the electromagnetic cloaks, an effect that is not possible in conventional absorbers of the same thickness, since the field energy is limited to straight line propagation. Similar reasoning explains the slightly improved scattering coefficient of the ideal `cloaking'' absorber compared to the reduced ones (Fig. \ref{scatteringcoeffabsorber}(e)). The reduced parameter sets allow imperfect field bending and, thus, some energy scatters off the inner core of the absorption coating \cite{mcguirk2008ctf}.

The proposed absorber designs also inherit the reflectionless property of the cloaks with both ideal and reduced parameter sets. The characteristic impedance of the absorbers at the outer boundary is $Z|_{r=R_{2}}=\sqrt{\frac{\mu_{0}\mu_{z}(1-\jmath\tan\delta)}{\varepsilon_{0}\varepsilon_{\phi}(1-\jmath\tan\delta)}}=\sqrt{\frac{\mu_{0}}{\varepsilon_{0}}}$, which is the free space wave impedance . As a result, the absorbers are matched to the surrounding free space and there is minimal backscattering. This is indeed observed in Fig. \ref{scatteringcoeffabsorber}(c), (d), where the field at the backscattering angle of $180^{\circ}$ remains low as the losses are being increased, which is advantageous for the efficient use of the devices.

Finally, we would like to address the weak omnidirectional scattering that was previously observed in Figs. \ref{scatteringcoeffabsorber}(c), (d) as the losses are increased, especially between the angles of $90^{\circ}$ to $120^{\circ}$ and $240^{\circ}$ to $270^{\circ}$. In order to explain this particular response of the device, the cylindrical wave expansion is used \cite{Ruan}. The total magnetic field inside and outside of the cylindrical 2-D ``cloaking'' absorber is given, respectively, by $H^{in}_{z}=\displaystyle\sum_{l}[a^{1}_{l}J_{l}(k_{1}(r-R_{1}))\exp(jl\phi)+a^{2}_{l}H_{l}(k_{1}(r-R_{1}))\exp(jl\phi)]$ and $H^{out}_{z}=\displaystyle\sum_{l}[a^{inc}_{l}J_{l}(k_{0}r)\exp(jl\phi)+a^{sc}_{l}H_{l}(k_{0}r)\exp(jl\phi)]$. Here $J_{l}$, $H_{l}$ are the $l^{th}$-order Bessel and Hankel functions of the first kind, respectively. In order to computer the fields, the expansion coefficients $a^{inc}_{l},a^{sc}_{l},a^{1}_{l},a^{2}_{l}$ need to be calculated by applying the proper boundary conditions at the inner and outer surfaces of the absorption coating. It was obtained in \cite{Ruan} that $a^{2}_{l}=0$ for both the ideal and reduced parameter sets, which also holds true in the proposed absorbers.

Nevertheless, for the scattering expansion coefficients in the current ``cloaking'' absorber $a^{sc}_{l}\neq a^{2}_{l}=0$. This is a direct result from the differences existing between the phase variations in the arguments of the Bessel and Hankel functions in the above expansions. More precisely, at the outer surface $r=R_{2}$ of the device, the phase variation of the total incident wave is $k_{0}R_{2}$, whilst the phase variation of the total field inside the absorber is equal to $k_{1}(R_{2}-R_{1})$. However, unlike the case of the ideal and reduced cylindrical lossless cloaks, these two quantities are not equal when losses are introduced. The phase variation of the field inside the absorber is: $k_{1}(R_{2}-R_{1})=\sqrt{\varepsilon_{0}\varepsilon_{\phi}(1-\jmath\tan\delta)\mu_{0}\mu_{z}(1-\jmath\tan\delta)}=k_{0}R_{2}(1-\jmath\tan\delta)$, which is directly derived from the ideal and reduced parameters. Meanwhile, the phase variation of the field outside the absorber is $k_{0}R_{2}=\sqrt{\varepsilon_{0}\mu_{0}}R_{2}$. This phase mismatch which occurs as the loss tangent is increased from zero, gives rise to the weak scattering observed at the sides of the absorber in Figs. \ref{scatteringcoeffabsorber}(c), (d). Nevertheless, this scattering is compensated by the bigger shadow achieved when more losses are introduced and the overall performance of the ``cloaking'' absorber is mostly unaffected by this slight imperfection.

The proposed discrete reduced set absorbing device can be constructed with ten concentric arrays of split-ring resonators or electric ring resonators printed on a dielectric substrate, as was firstly demonstrated in \cite{Schurig}. The varying collision frequency values of the ``cloaking'' absorber can be controlled with different practical techniques. For the microwave frequency regime, where the dielectric losses are dominant \cite{raynolds2003olf}, different substrate dielectric materials can be chosen, or lossy nanoparticles can be introduced into the substrate. For higher frequencies, such as infrared, the ohmic losses are more significant \cite{raynolds2003olf}, which leads to the solution of changing the metallic material of the resonator particles (for example, gold is more lossy than silver at infrared frequencies) or varying their dimensions.

To conclude, a set of novel metamaterial absorbers is proposed based on the manipulation of losses inside electromagnetic cloaks, yielding absorptivity values larger than $80\%$. It is explained that a discrete reduced ``cloaking'' absorber can be constructed with existing metamaterial technologies. These devices could be used as narrow-band bolometers, due to their dispersive nature; they also exhibit a moderate wideband response, which is ideal for RCS reduction. Another useful application is the total isolation of an antenna placed inside the absorber from low-frequency surrounding noise. It is demonstrated that the devices are reflectionless and efficient in terms of absorption. Moreover, they have subwavelength thickness, which is highly desirable for absorbers operating in VHF or UHF bands. Finally, such absorbers consist one more useful addition to the overlooked impact of metamaterials to absorbing technology.

\bibliography{referencesall}

\end{document}